# Microwave Second-Harmonic Response of Ceramic $MgB_2$ Samples

A. Agliolo Gallitto, G. Bonsignore and M. Li Vigni

Dipartimento di Scienze Fisiche ed Astronomiche, Università di Palermo, via Archirafi 36, 90123 Palermo, Italy

**Abstract**

Nonlinear microwave response of different ceramic $MgB_2$ samples has been investigated by the technique of second-harmonic emission. The second-harmonic signal has been investigated as a function of temperature, DC magnetic field and input microwave power. The attention has mainly been devoted to the response at low magnetic fields, where nonlinear processes arising from motion of Abrikosov fluxons are ineffective. The results show that different mechanisms are responsible for the nonlinear response in the different ranges of temperature. At low temperatures, the nonlinear response is due to processes involving weak links. At temperatures close to $T_c$, a further contribution to the harmonic emission is present; it can be ascribed to modulation of the order parameter by the microwave field and gives rise to a peak in the temperature dependence of the harmonic signal.

**Corresponding author**:

Dott. Aurelio AGLIOLO GALLITTO
Dip.to di Scienze Fisiche ed Astronomiche, via Archirafi 36, I-90123 Palermo (Italy)
Tel: +39 091 6234207; Fax: +39 091 6162461; E-mail: agliolo@fisica.unipa.it

# 1. Introduction

Since the discovery of the superconductivity at 40 K in magnesium diboride [1], a large number of studies have been devoted to understanding the mechanism of superconductivity in $MgB_2$ as well as to explore its potential for technological applications [2-4]. Indeed, the malleability and ductility of the compound, due to its metallic nature, make $MgB_2$ a promising candidate for producing wires, tapes and layers even of wide areas, to be used for a large variety of devices. An important property of $MgB_2$ is that the grain boundaries do not appreciably affect the transport properties of this material, in contrast to what occurs in cuprate superconductors [2]. This property, very important for applications, has been highlighted by different authors; it seems to be independent of the preparation method of the material.

One of the technological fields in which the use of superconducting materials is particularly convenient concerns the implementation of superconductor-based microwave (mw) devices [3-6]. On the other hand, it has been widely shown that high-$T_c$ superconductors are characterized by markedly nonlinear properties when exposed to intense *em* fields up to microwave frequencies [5-16]. Several mechanisms responsible for the nonlinear response have been recognized, whose effectiveness depends on the temperature, magnetic field and type of superconductors. In particular, it has been shown that at low magnetic fields and low temperatures the main source of nonlinearity arises from processes occurring in weak links [9, 14, 17-19]. The nonlinearity is really the main limiting factor for application of superconductors in passive microwave devices [5, 6]; on the other hand, it can be conveniently exploited for assembling active mw devices [5]. For this reason, it is of great relevance recognizing the mechanisms responsible for the nonlinear response and determining the conditions in which nonlinear effects are important.

A suitable method for investigating the nonlinear microwave response consists in detecting signals oscillating at the harmonic frequencies of the driving field [7-14]. In this paper, we report results of second-harmonic (SH) emission by ceramic $MgB_2$ samples, prepared by different methods. The signal radiated at the SH frequency of the driving field has been investigated as a function of temperature and external magnetic field. Particular attention has been devoted to the investigation of the properties of the SH signal for applied magnetic fields lower than $H_{c1}$, when nonlinear processes arising from motion of Abrikosov fluxons are not effective. The experimental results are discussed in the framework of models reported in the literature. Though grain boundaries in $MgB_2$ do not appreciably affect the transport properties, we show that, even in this superconductor, the main source of SH emission at low magnetic fields and low temperatures is related to the presence of weak links.

## 2. Experimental apparatus and samples

We report experimental results of SH emission in Alfa Aesar $MgB_2$ powder and a bulk $MgB_2$ sample. The bulk sample has been synthesized by the direct reaction of boron powder with a lump of magnesium metal, both of purity better than 99.95% [20].

Fig. 1 shows the real component of the AC susceptibility at 100 kHz as a function of the temperature for 5 mg of Alfa Aesar powder, which we indicate as sample $P_\alpha$, and the bulk sample, which we indicate as B. Although the samples have roughly the same transition temperature, their quality is different. In particular, $P_\alpha$ has very likely an inhomogeneous superconducting transition with two characteristic temperatures about 1.5 K apart, while the bulk sample shows a single-phase transition of width of about 1 K.

The sample is located inside a bimodal rectangular cavity resonating at the two angular frequencies $\omega$ and $2\omega$, in a region where the magnetic fields $H(\omega)$ and $H(2\omega)$ are parallel and of maximal intensity. The fundamental frequency $\omega/2\pi$ is $\approx$ 3 GHz. The $\omega$-mode of the cavity is fed by a pulsed mw field (pulse repetition rate 200 pps, pulse width 5 μs, maximum peak power ≈ 50 W). The sample is also exposed to a static magnetic field, $H_0$, which can be varied from 0 to 10 kOe. The harmonic signals are detected by a superheterodyne receiver. Further details of the experimental apparatus are reported in Ref. [12]. The intensity of the SH signal has been measured as a function of the external magnetic field, the temperature and the input power. All the measurements have been performed with ***H**($\omega$)*||**H**(2$\omega$)*||**H**$_0$*.

### 3. Experimental results

Fig. 2 shows the SH signal intensity as a function of the temperature for $P_\alpha$ and B samples at $H_0$ = 10 Oe. Though the SH signal has different intensity in the two samples, its temperature dependence is similar in both samples; in particular, the SH emission is significant in the whole range of temperatures investigated and exhibits an enhanced peak at temperatures near $T_c$. The curve relative to the $P_\alpha$ sample exhibits a kink at $T \approx 38$ K which is ascribable to the inhomogeneity of the superconducting transition.

Measurements of SH emission have been performed in other ceramic $MgB_2$ samples prepared by different methods; the SH signal shows roughly the same peculiarities, except for the intensity. For this reason, in the following we report the results of SH emission, obtained on varying the DC magnetic field and the input power, in the $P_\alpha$ sample. Indeed, this sample exhibits an intense SH signal that can be investigated in a wide range of input power.

Fig. 3 show the temperature dependence of the SH signal observed in $P_\alpha$ at different input power levels. As one can see, except for a variation of the SH signal intensity, the peculiarities of the SH-vs-$T$ curves do not change with the input power; in particular, the temperature range at which the near-$T_c$ peak takes place is independent of the input power level.

In Fig. 4 we report the SH signal intensity as a function of the DC magnetic field for the $P_\alpha$ sample, at $T = 4.2$ K. Full symbols refer to the results obtained in the zero-field-cooled (ZFC) sample on increasing the field for the first time. As expected from symmetry considerations, the SH signal is zero at $H_0 = 0$ (the noise level is ~ – 80 dBm); on increasing the field it abruptly increases, exhibits a maximum at $\approx 3$ Oe, decreases monotonically up to ~ 40 Oe and then slowly increases until it reaches a value that remains roughly constant up to high fields. Open symbols describe the field dependence of the SH signal observed on decreasing $H_0$, after the first run to high fields. It is worth noting that the low-field structure disappears irreversibly after the sample has been exposed to high fields; in this case, even at $H_0 = 0$ the intensity of the SH signal takes on roughly the same value as the one measured at high fields.

Fig. 5 shows the low-field behavior of the SH signal, observed in the ZFC $P_\alpha$ sample by cycling the magnetic field in the range $-H_{max} \div +H_{max}$. The signal shows a hysteretic loop having a butterfly-like shape; on increasing the value of $H_{max}$, the hysteresis is more and more enhanced and the sharp minima, observed at low fields, move away from each other. The hysteretic loop maintains the same shape in subsequent field runs as long as the value of $H_{max}$ is not changed. It is worth remarking that no hysteresis is observed when $H_{max}$ is smaller than the DC field at which the maximum in the SH-vs-$H_0$ curves occurs; furthermore, the low-field structure disappears irreversibly for $H_{max}$ of the order of 100 Oe.

Fig. 6 shows the field dependence of the SH signal in the $P_\alpha$ sample, at different input power levels. As one can see, the amplitude of the hysteresis increases on decreasing the input power, suggesting that the power dependence of the SH signal is different for increasing and decreasing fields. The inset shows the intensity of the SH signal as a function of the input power level, at $H_0 = 20$ Oe: full symbols show the results obtained at $H_0 = 20$ Oe reached at increasing fields; open symbols those obtained at $H_0 = 20$ Oe reached at decreasing fields.

Measurements performed in the bulk sample B have shown that the low-field SH signal exhibits qualitatively the same peculiarities of the signal detected in $P_\alpha$. Only the value of the DC field at which the maximum of the SH-vs-$H_0$ curve falls and the distance between the sharp minima, obtained cycling $H_0$ in the range $-H_{max} \div +H_{max}$, slightly differ in the two samples.

The low-field dependence of the SH signal has been investigated at different values of the temperature. We have found that the low-field hysteresis loop maintains roughly the same shape up to about 3 K below $T_c$. At temperatures closer to $T_c$, the peculiarities of the low-field SH signal change; in particular, the hysteretic behavior appears after the DC magnetic field has reached values of the order of 10 Oe. Fig. 7 shows the field dependence of the SH signal in the ZFC $P_\alpha$ sample at $T = 36.8$ K, obtained by varying $H_0$ up to the maximal value of 10 Oe. The dashed line shows the first run starting from $H_0 = 0$. The inset shows the field dependence of the SH signal obtained by cycling $H_0$ in the range $-2 \div +2$ Oe. A comparison between Fig. 7 and Fig. 5 shows that the hysteretic loop exhibits different shape at low and high temperatures. Furthermore, we would remark that at temperatures close to $T_c$ the amplitude of the hysteresis is independent of the input power level, in contrast with the results of Fig. 6. These findings suggest that the hysteretic behavior visible in Fig. 7 and that of Fig. 5 have different origin. On the other hand, it is reasonable to hypothesize that at temperatures near $T_c$ the sample goes into the mixed state for magnetic fields of the order of 10 Oe. So, the hysteresis loop of Fig. 7 can be ascribed to the critical state of intra-grain fluxons.

## 3. Discussion

The nonlinear response of high-$T_c$ superconductors to electromagnetic fields has been extensively investigated [7-19]. At low temperatures, it has been ascribed to extrinsic properties of the superconductors such as impurities, weak links [9, 11, 13-15, 17-19] or fluxon motion [7, 8, 11, 13]. On the contrary, nonlinearity at temperatures close to $T_c$ can be related to intrinsic properties of the superconducting state [10-14]; in particular, it has been ascribed to modulation of the order parameter induced by the mw field. Our results on $MgB_2$ suggest that, also in this class of superconductors, different mechanisms come into play in the SH emission. In the following, we discuss the behavior of SH emission at low and high temperatures, by considering models reported in the literature.

*3.1 Nonlinear response at low temperatures*

Harmonic generation has been thoroughly investigated in both ceramic and crystalline YBaCuO samples [7-13, 17-19]. It has been shown that the low-field harmonic emission by high-quality crystals is noticeable only at temperatures near $T_c$ [10, 12, 13], while in ceramic samples it is significant also at low temperatures [9, 17]. This finding has pointed out that the low-field and low-$T$ harmonic signals are due to nonlinear processes occurring in weak links.

In the presence of weak links, two nonlinear processes may come into play. Harmonic emission is expected when supercurrents are induced, by the $H_0$ and $H(\omega)$ fields, in loops containing Josephson junctions (JJ). In this case, the harmonic emission is strictly related to the intrinsic nonlinearity of the Josephson current [9, 17]. On the other hand, inter-grain dynamics of Josephson fluxons (JF) in the critical state may give rise to harmonic emission [18, 19]. None of the models elaborated to describe these nonlinear processes accounts for the hysteretic loops of the low-field SH signal. On the other hand, results similar to those of Figs. 5 and 6 have been obtained in BaKBiO crystals [14]; they have been justified by supposing that the above-mentioned nonlinear processes involving weak links come into play simultaneously. We have shown that the combined effect of the $2\omega$ magnetization arising from the supercurrent loops containing JJ and that arising from the dynamics of JF can justify the peculiarities of the hysteretic behavior of the low-field SH signal [14]. According to what discussed in Ref. [14], we suggest that also in ceramic $MgB_2$ samples the low-$T$ and low-field SH signal arises from nonlinear processes due to the presence of weak links. This hypothesis is supported by the irreversible loss of the low-field structure, observed in ZFC samples, after the samples had been exposed to high fields (see Fig. 4). Indeed, when $H_0$ reaches values higher than $H_{c1}$, Abrikosov fluxons penetrate the grains and the JJ are decoupled by the applied field and/or the trapped flux.

Concerning the origin of the SH emission at magnetic fields higher than $H_{c1}$, it has been shown that even harmonics can arise in superconductors in the critical state exposed to high-frequency magnetic fields [8]. In this case, because of the rigidity of the fluxon lattice the superconducting sample operates a rectification process of the AC field. Ciccarello et al. [8] have elaborated a model concerning these effects. It has been shown that a peculiarity of the SH signal of superconductors in the critical state is the presence of enhanced dips in the SH-vs-$H_0$ curves, after the inversion of the magnetic-field-sweep direction, independently of the $H_0$ value at which the inversion is operated. The dips can be ascribed to the fact that, after the inversion of the magnetic-field-sweep direction, a certain variation of the DC field is necessary in order to establish a reverse critical state. The presence of the enhanced dip in the SH-vs-$H_0$ curve after the inversion of the magnetic-field sweep, visible in Fig. 4 at the extreme of the plot, indicates that also in $MgB_2$, when the sample is in the mixed state, SH emission is due to the intra-grain fluxon dynamics. We suggest that this mechanism is responsible also for the hysteretic behavior of the SH signal of Fig. 7, in agreement with Ref. [13].

*3.2 Nonlinear response at temperatures near $T_c$*

Several studies carried out in conventional as well as high-$T_c$ superconductors have suggested that the enhanced harmonic emission at temperatures close to $T_c$ can be ascribed to the time variation of the order parameter induced by the *em* field [10-14]. In particular, the features of second- and third-harmonic signals detected in high-$T_c$ superconductors have been accounted for in the framework of the two-fluid model with the additional hypothesis that the *em* field, which penetrates in the surface layers of the sample, weakly perturbs the partial concentrations of the normal and condensate fluids [10, 12, 14]. In this case, it is expected that the intensity of the SH signal is proportional to the effective surface of the sample, through which the *em* field penetrates.

The experimental results reported in Fig. 2 show that the temperature dependence of the SH signal is similar in the two samples, apart from the signal intensity. This finding would suggest that the SH emission in $MgB_2$ arises from the same mechanism in all the range of temperatures investigated; however, some experimental results disagree with this hypothesis.

In the framework of the models that consider current loops containing JJ it is expected a $2\omega$ magnetization proportional to the Josephson critical current [9, 17] that, as well known, decreases monotonically on increasing the temperature; so, this process cannot account for the enhanced peak detected near $T_c$. On the contrary, Ji *et al*. [18] have suggested that harmonic signals due to dynamics of inter-grain JF in the critical state can exhibit a near-$T_c$ peak. In the framework of their model, the intensity of the harmonic signals strongly depends on the $H_J^*/H_{AC}$ ratio, where $H_J^*$ is the full penetration field in the inter-grain region. By taking into account the temperature dependence of $H_J^*$, it is possible to show that a near-$T_c$ peak in the harmonic emission is expected whenever at low temperatures $H_J^*$ is larger than $H_{AC}$. The authors of Ref. [18], measuring third-harmonic emission as a function of temperature in ceramic YBaCuO, have detected a near-$T_c$ peak in a bulk sample and have not detected any peak in a powdered sample. They ascribe this finding to the fact that in the bulk sample at low temperatures $H_J^* > H_{AC}$, while in the powdered sample, due to the reduced grain sizes, $H_J^* < H_{AC}$.

Although the authors of Ref. [18] do not discuss the temperature dependence of SH emission, but only third harmonic, from their model it is possible to show that also the SH signal can exhibit a near-$T_c$ peak. However, in the framework of this model, the value of the temperature at which the maximum of the near-$T_c$ peak falls, as well as the peak width, strongly depends on the $H_J^*/H_{AC}$ ratio. Our results show that the width and the position of the near-$T_c$ peak are nearly the same in the two samples (Fig. 2) and they are roughly independent of the input power level (Fig. 3), suggesting that the enhanced SH emission we observe at temperatures close to $T_c$ is not due to the motion of JF in the critical state.

In order to corroborate the hypothesis that the low-$T$ and high-$T$ SH signals originate from two different mechanisms, the former from processes involving weak links and the latter from modulation of the order parameter, we have performed measurements in other two samples, which we indicate as $P_{\alpha c}$ and $P_{\alpha d}$, obtained by handling 5 mg of Alfa Aesar powder. $P_{\alpha c}$ has been obtained by crushing $\approx$ 5 mg of pristine Alfa Aesar powder; $P_{\alpha d}$ has been obtained by dispersing the crushed powder in polystyrene, with a ratio of ~1:10 in volume. The grinding process of the pristine powder has been performed to reduce the grain size; consequently, $H_J^*$ in $P_{\alpha c}$ should be smaller than in $P_\alpha$. The dispersion of the grains of $P_{\alpha c}$ in polystyrene has been performed to reduce the number of weak links, maintaining unchanged the effective surface of the sample. The temperature dependence of the SH signal obtained in the three samples, $P_\alpha$, $P_{\alpha c}$ and $P_{\alpha d}$, is shown in Fig. 8. For the sake of clearness, the three curves have been shifted one with respect to the other, as shown in the figure. These results show that the main peculiarities of the SH signal observed in $P_{\alpha c}$ and $P_{\alpha d}$ are the same of those obtained in the $P_\alpha$ sample, except for the relative intensity of the near-$T_c$ peak and the low-$T$ signal.

From Fig. 8, one can see that the ratio of the intensity of the near-$T_c$ peak and the signal at the lowest temperature investigated is about 7 dB for $P_\alpha$ and 12 dB for $P_c$, this finding could be related to the combined effect of different factors, such as variations of the number of weak links, the mean size of inter-grain regions and the effective surface of the sample. The finding that the value of temperature at which the maximum of the near-$T_c$ peak falls is the same in the two samples corroborates the hypothesis that the enhanced peak observed at temperatures near $T_c$ is not ascribable to motion of JF in the critical state.

A comparison between the SH-vs-$T$ curves of the samples $P_{\alpha c}$ and $P_{\alpha d}$ of Fig. 8 shows that the intensity of the near-$T_c$ peak is the same for the two samples, while a further decrease of the low-$T$ signal is observed in $P_{\alpha d}$, i.e. after dispersing the powder in polystyrene. These results suggest that the low-$T$ and the near-$T_c$ SH emission have different origin. In particular, the decrease of the SH signal intensity at low temperatures confirms that the low-$T$ signal arises from processes involving weak links. On the contrary, the unchanged intensity of the near-$T_c$ peak strongly suggests that, also in $MgB_2$, the enhanced nonlinearity at temperatures close to $T_c$ may arise from modulation of the order parameter by the *em* field.

## 4. Conclusion

We have reported experimental results on the microwave second-harmonic response of ceramic $MgB_2$. The SH signal has been investigated as function of the temperature, the input power level and the DC magnetic field. The results show that, similarly to what occurs in other high-$T_c$ superconductors, several mechanisms are responsible for the nonlinear microwave response of $MgB_2$; their effectiveness depends on the temperature and the intensity of the external magnetic field. The results obtained at low temperatures have shown that, although the presence of weak links in $MgB_2$ does not noticeably affect the transport properties, it is the main source of nonlinearity at low magnetic fields and low temperatures. After exposing the sample to magnetic fields higher than $H_{c1}$, weak links are decoupled and the SH emission originates from dynamics of intra-grain fluxons in the critical state. At temperatures close to $T_c$, a peak in the temperature dependence of the SH signal is present; it may arise from modulation of the order parameter by the microwave field. The near-$T_c$ peak detected in $MgB_2$ is wider than that observed in YBaCuO crystals [10, 12], this can be due to contribution of the SH signal arising from processes occurring in weak links as well as to the inhomogeneous broadening of the superconducting transition. Further studies of high-quality $MgB_2$ samples are necessary to better investigate the harmonic emission near $T_c$ of the $MgB_2$ superconductor.

*Acknowledgements*

The authors are very glad to thank G. I. Leviev and D. V. Shovkun for helpful suggestions, G. Lapis and G. Napoli for technical assistance, N. N. Kolesnikov and M. P. Kulakov for supplying the bulk $MgB_2$ sample. Work performed in the framework of the collaboration between the University of Palermo and the ISSP of the Russian Academy of Sciences (Grant Coll. Int. Li Vigni 2002, University of Palermo).

**Figure captions**

Fig. 1 Temperature dependence of the real component of the AC susceptibility at 100 kHz, in $P_\alpha$ and B samples, normalized to the values obtained at $T = 4.2$ K.

Fig. 2 SH signal intensity as a function of the temperature for samples $P_\alpha$ and B. DC magnetic field $H_0 = 10$ Oe; input peak power $P_{in} \approx 30$ dBm.

Fig. 3 SH signal intensity as a function of the temperature for $P_\alpha$, at three different input power levels.

Fig. 4 SH signal intensity as a function of the DC magnetic field for $P_\alpha$. Full symbols refer to the results obtained in the ZFC sample on increasing the field for the first time. Open symbols describe the field dependence of the SH signal on decreasing $H_0$ after the first run to high fields. Input peak power $\approx 27$ dBm. Lines are leads for eyes.

Fig. 5 Low-field behavior of the SH signal, observed in the ZFC $P_\alpha$ sample by cycling the magnetic field in range $-H_{max} \div +H_{max}$. $T = 4.2$ K; input peak power $\approx 30$ dBm. The three curves refer to different values of $H_{max}$. Lines are leads for eyes.

Fig. 6 Magnetic field dependence of the SH signal, obtained in $P_\alpha$ sample by cycling $H_0$ in the range $-35 \div +35$ Oe, at three different values of the input power. The inset shows the SH signal intensity as a function of the input power level at $H_0 = 20$ Oe, reached at increasing fields (full symbols) and at decreasing fields (open symbols). $T = 4.2$ K. Lines are leads for eyes.

Fig. 7 Magnetic field dependence of the SH signal observed in the ZFC $P_\alpha$ sample at $T = 36.8$ K. The dashed line refers to the results obtained on increasing the field for the first time. Lines are leads for eyes.

Fig. 8 SH signal intensity as a function of the temperature in the samples $P_\alpha$, $P_{\alpha c}$ and $P_{\alpha d}$, as displayed from the top. $H_0$ = 10 Oe; input peak power ≈ 30 dBm. The curves relative to $P_\alpha$ and $P_{\alpha c}$ samples have been shifted upwards of 20 dB and 10 dB, respectively.

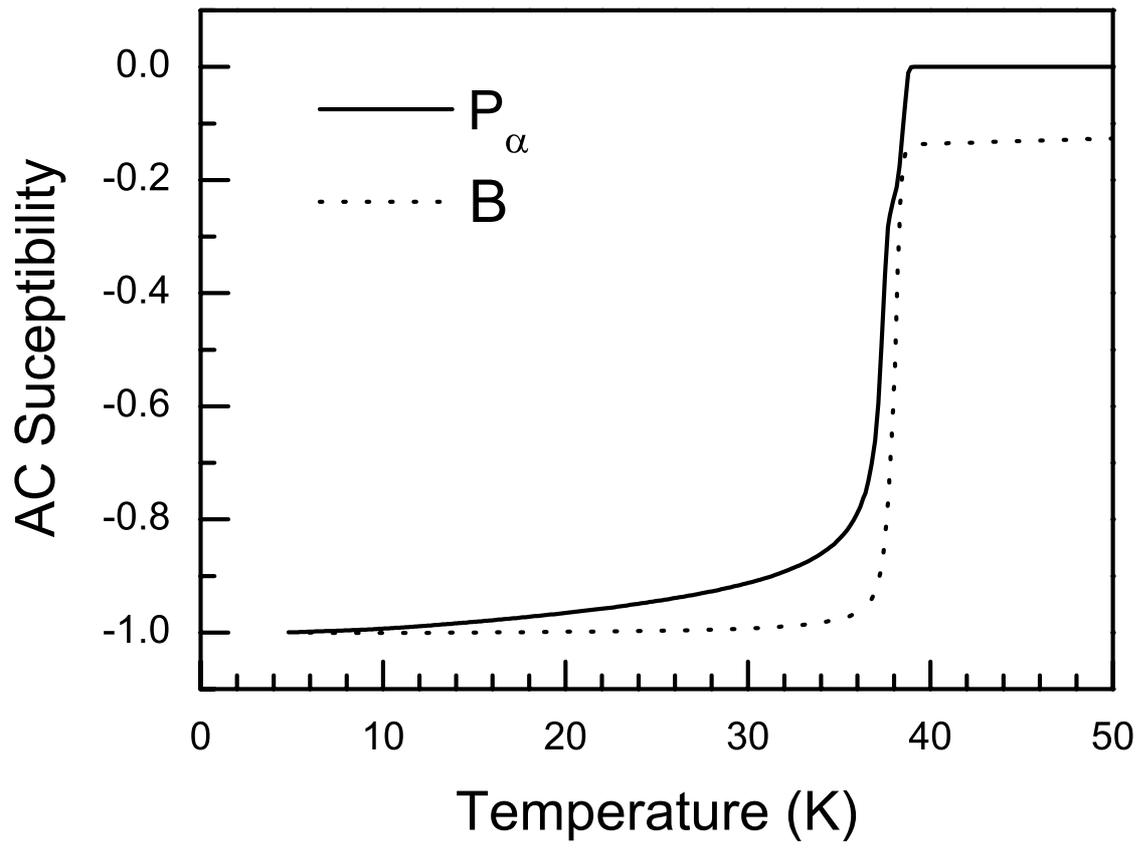

Fig.1

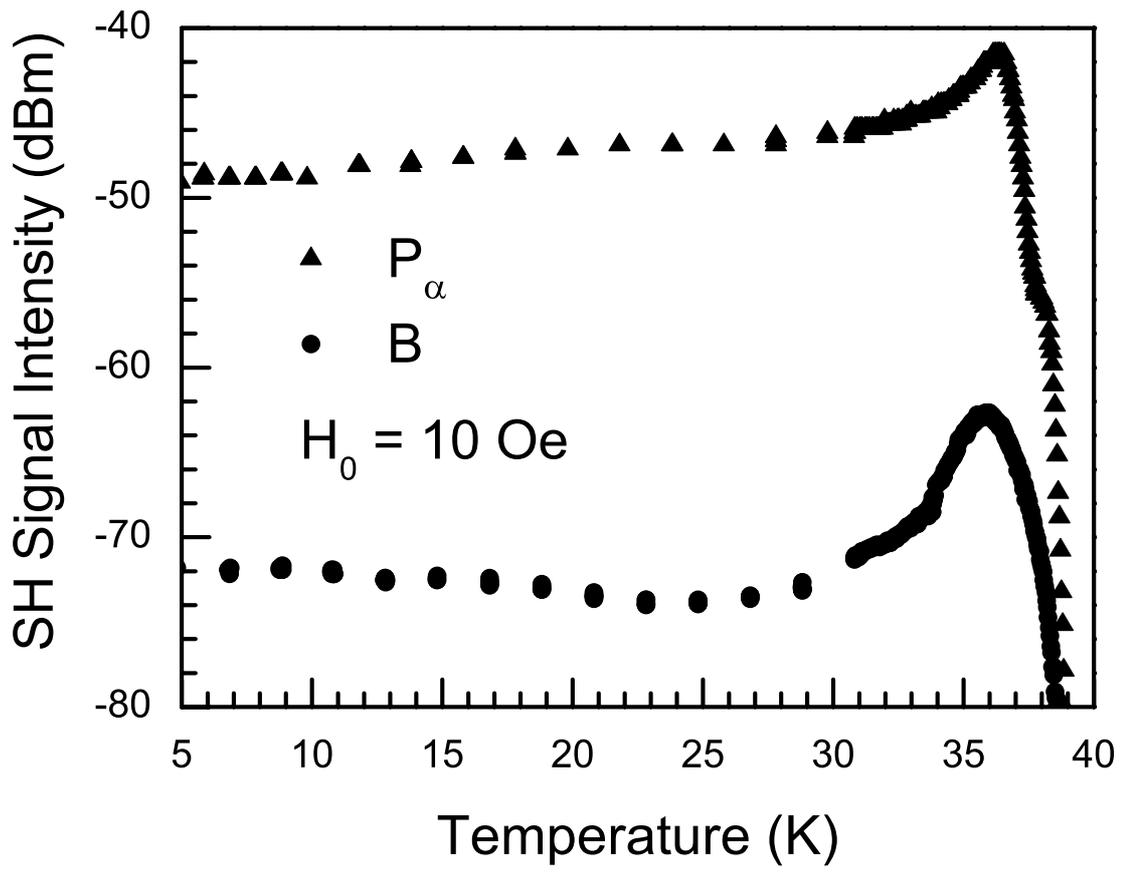

Fig. 2

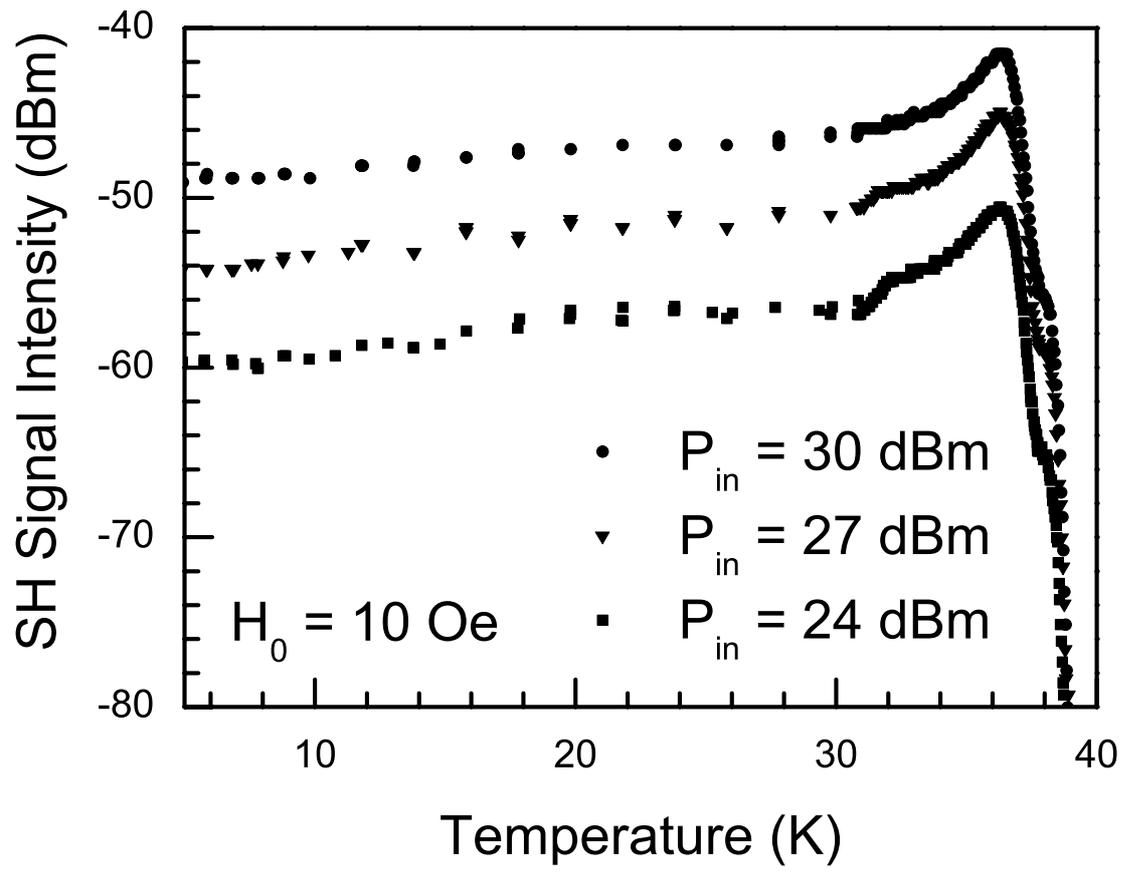

**Fig.3**

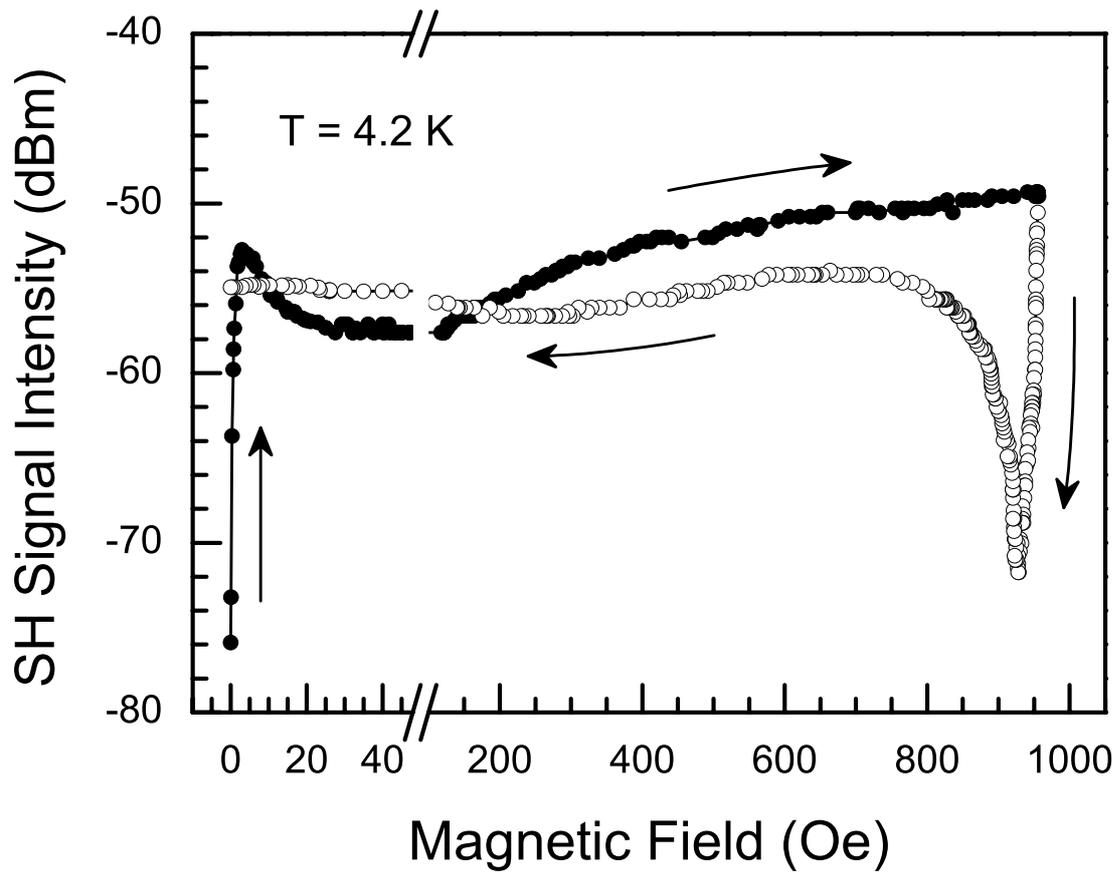

Fig.4

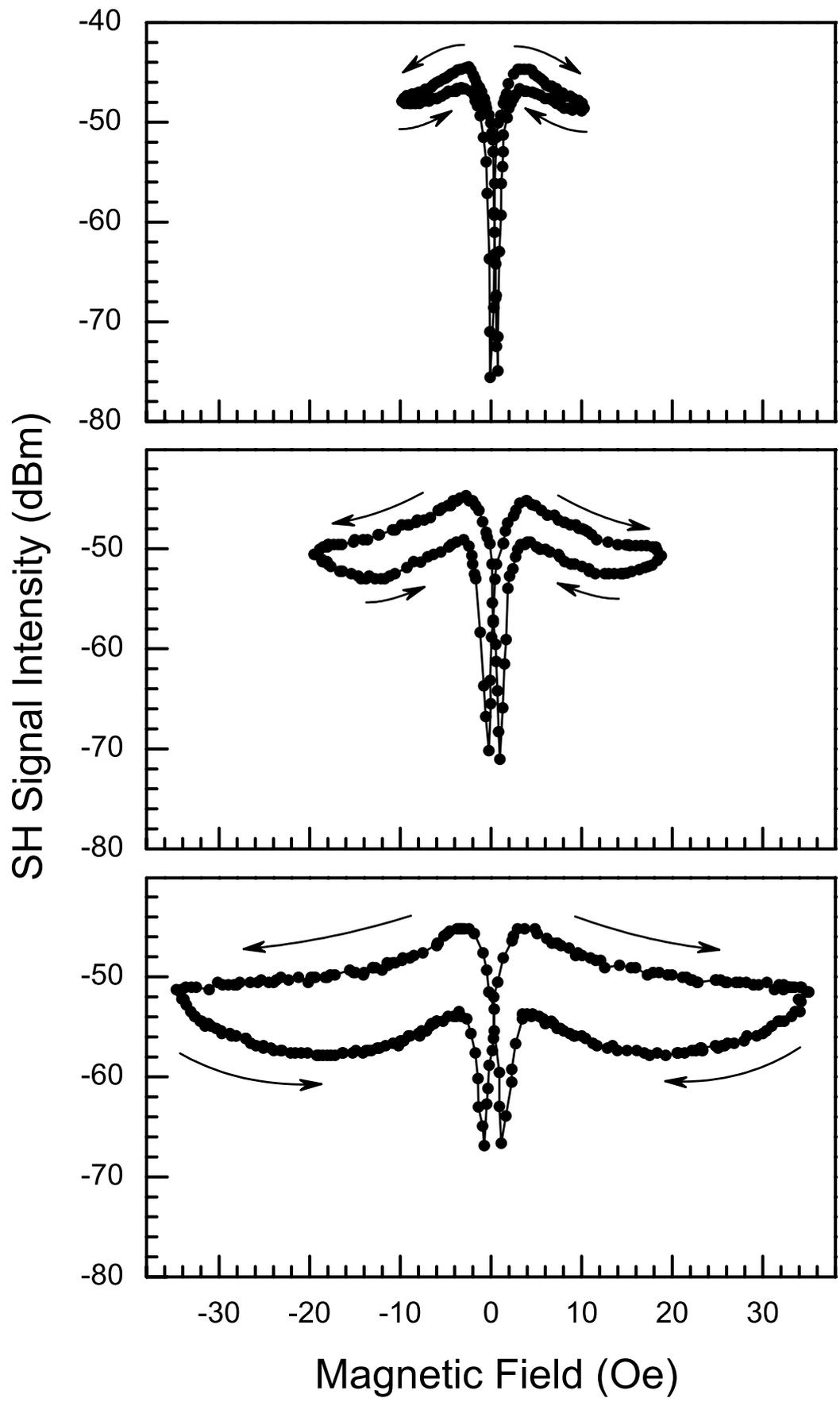

Fig.5

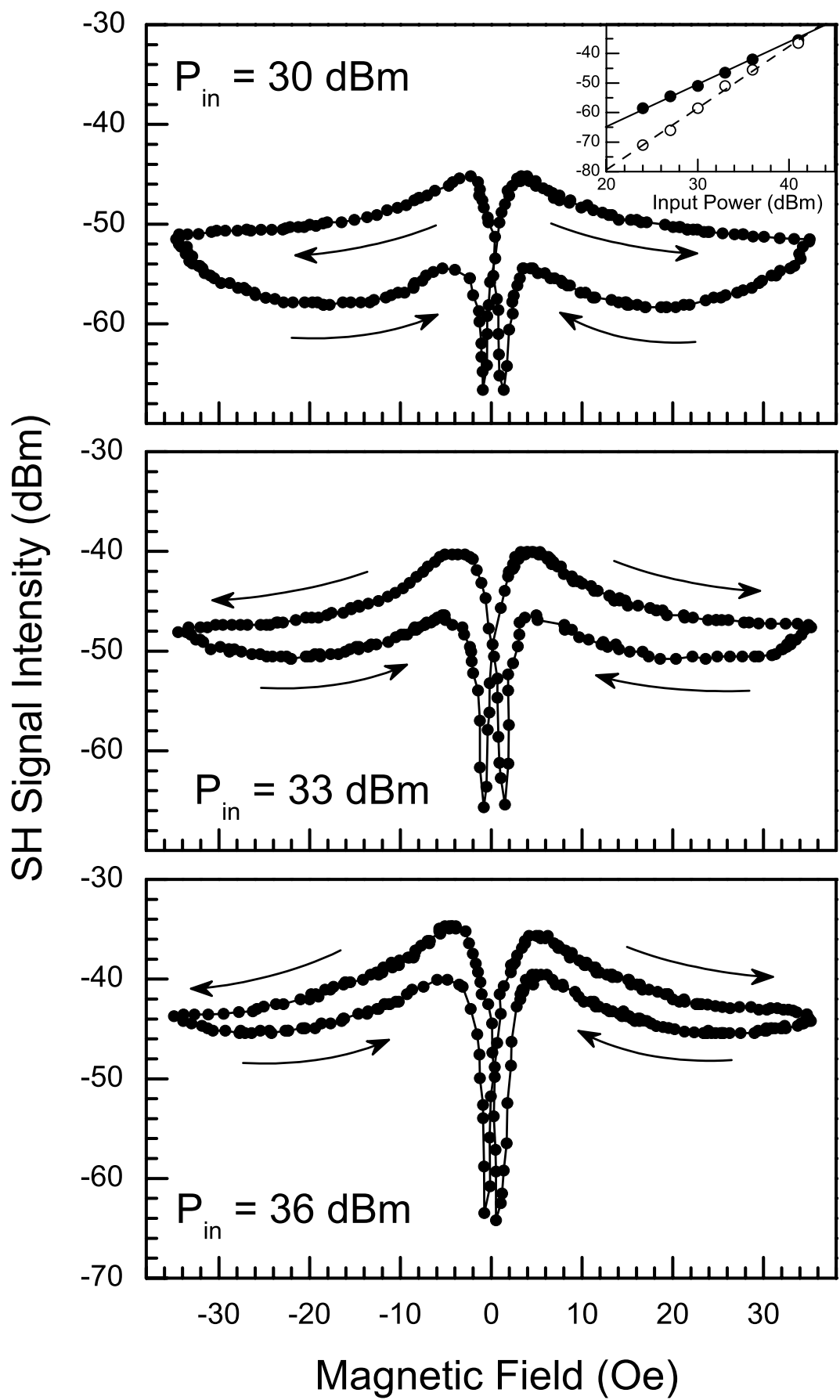

Fig. 6

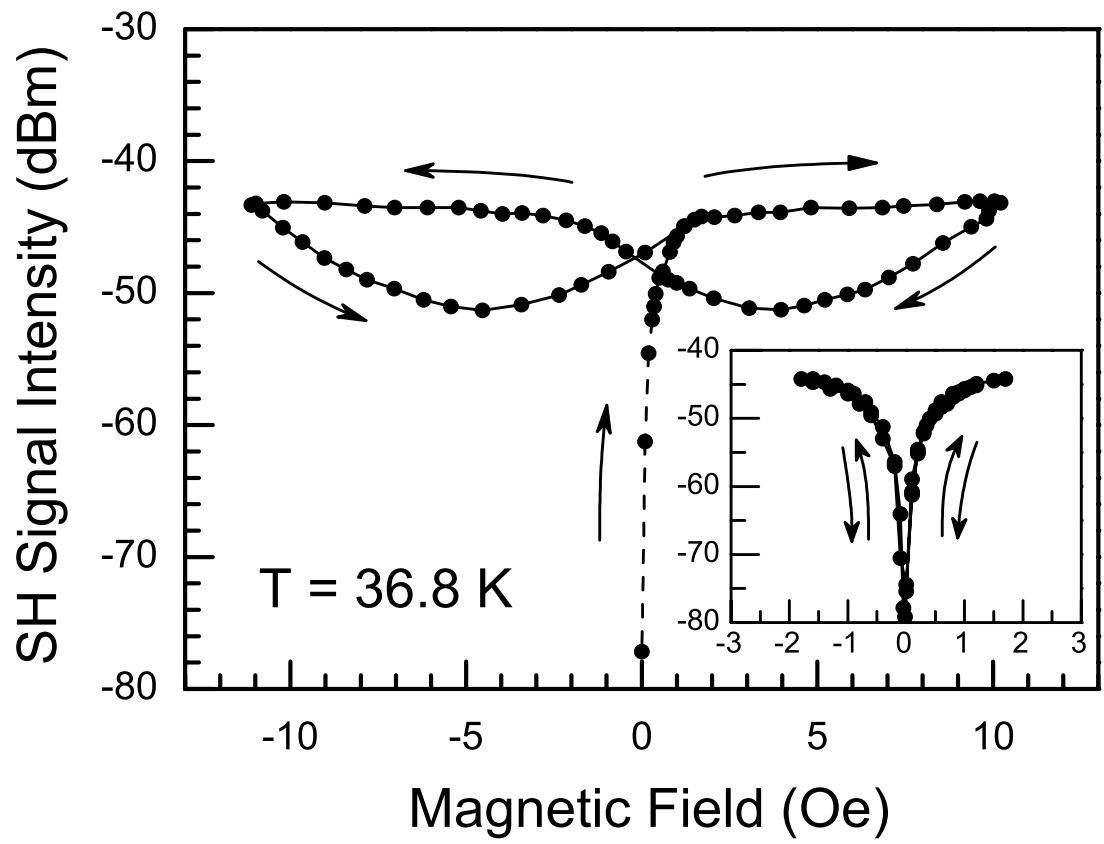

Fig.7

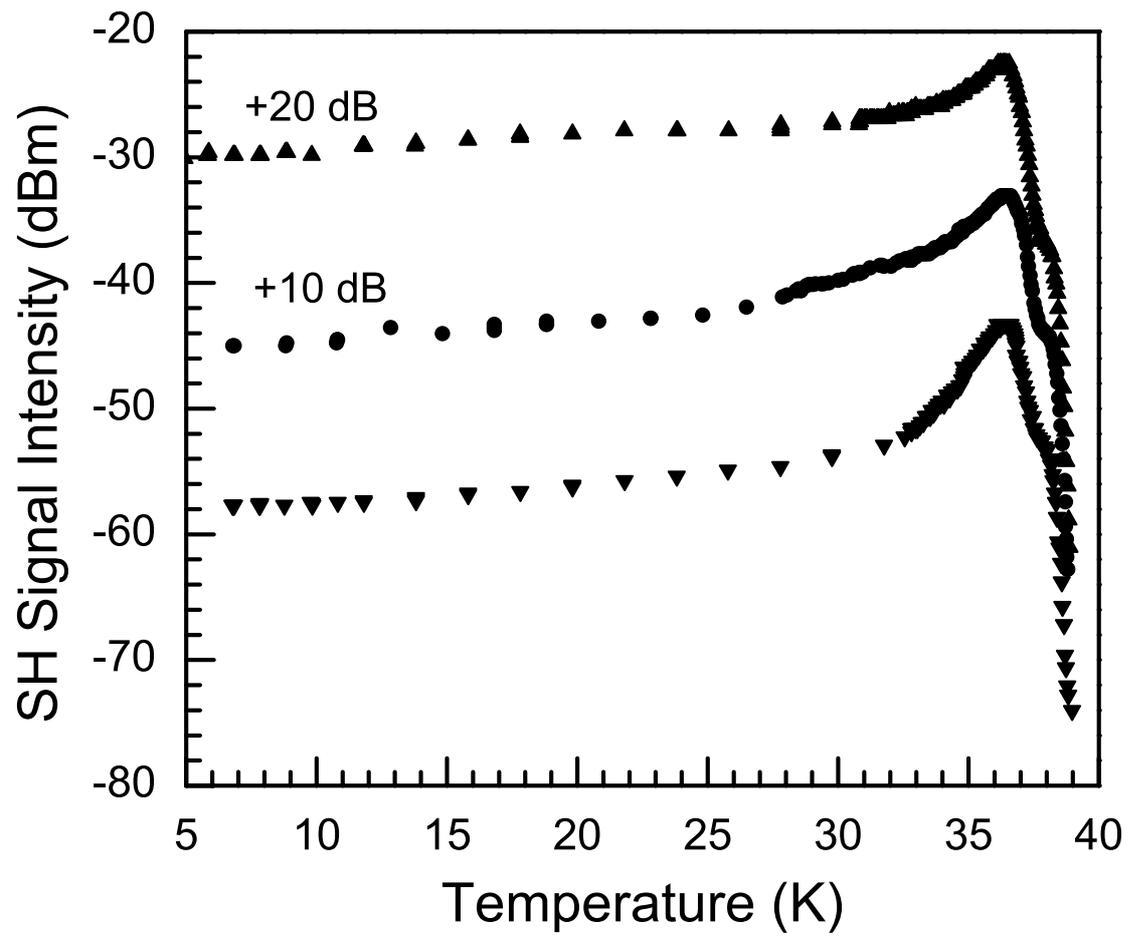

Fig.8